\documentclass[aps,prc,twocolumn,showpacs,floatfix,nofootinbib]{revtex4}
\usepackage{graphicx} 
\usepackage{dcolumn} 
\usepackage{bm}         
\begin{document}

\title{Kelvin-Helmholtz instability in high-energy heavy-ion collisions}

\author{
 L.P. Csernai$^{1,2,3}$, D.D. Strottman$^{2,3}$, and Cs. Anderlik$^4$}

\affiliation{
$^1$
Department of Physics and Technology,
University of Bergen, 5007, Bergen, Norway\\
$^2$
MTA Wigner Research Centre for Physics,
1525 Budapest, Pf. 49, Hungary\\
$^3$
Frankfurt Institute for Advanced Studies,
Johann Wolfgang Goethe University\\
Ruth-Moufang-Str. 1, 60438 Frankfurt am Main, Germany\\
$^4$
Uni Computing, Thorm{\o}hlensgate 55, N-5008 Bergen, Norway
}
\date{\today}

\begin{abstract}
The dynamical development of collective flow is studied in a  
(3+1)D fluid dynamical model, with globally symmetric, peripheral initial
conditions, which take into account the shear flow 
caused by the forward motion on the projectile side and the backward
motion on the target side. While at $\sqrt{s_{NN}} = 2.76A$\,TeV
semi-peripheral Pb+Pb collisions
the earlier predicted rotation effect is visible, at more peripheral
collisions, with high resolution and low numerical viscosity the
initial development of a Kelvin-Helmholtz instability is observed,
which alters the flow pattern considerably. This effect provides a
precision tool for studying the low viscosity of Quark-gluon Plasma. 
\end{abstract}


\maketitle

\section{ Introduction }

Global collective observables are becoming the most essential
in ultra-relativistic heavy ion reactions 
\cite{ALICE-Flow1}.
When we want to extract information
from experiments, both on the equation of state
(EoS) and the transport properties of matter \cite{Son,CKM},
we have to invoke a realistic description with a fully (3+1)D dynamical
evolution at all stages of the reaction, including the initial state. 

It is important to note that the phase transition to 
quark-gluon plasma (QGP) and consequent fluctuations may enhance 
the collective behavior of the system \cite{CK92}.
For the fluid dynamical (FD) initial state
we must have a system that is close to local
equilibrium; thus, at high energies the transition to QGP has
to happen earlier than the formation of the locally
equilibrated initial state.

The (3+1)D, relativistic FD model we use to describe energetic
heavy ion reactions is well established and describes the measured
collective flow reliably \cite{hydro1,hydro2}. We use the
Particle in Cell (PIC) method in which an Eulerian grid contains a 
very large number of Lagrangian  marker 
particles that move with the matter. This method enables us 
to follow the motion of the fluid 
with good precision. At Large Hadron Collider (LHC) energies 
in these calculations we observed a
significant rotation of the QGP fluid in peripheral
collisions, which leads to observable consequences \cite{hydro2}.

\begin{figure}[h]
 \centering
 \includegraphics[width=3.4in]{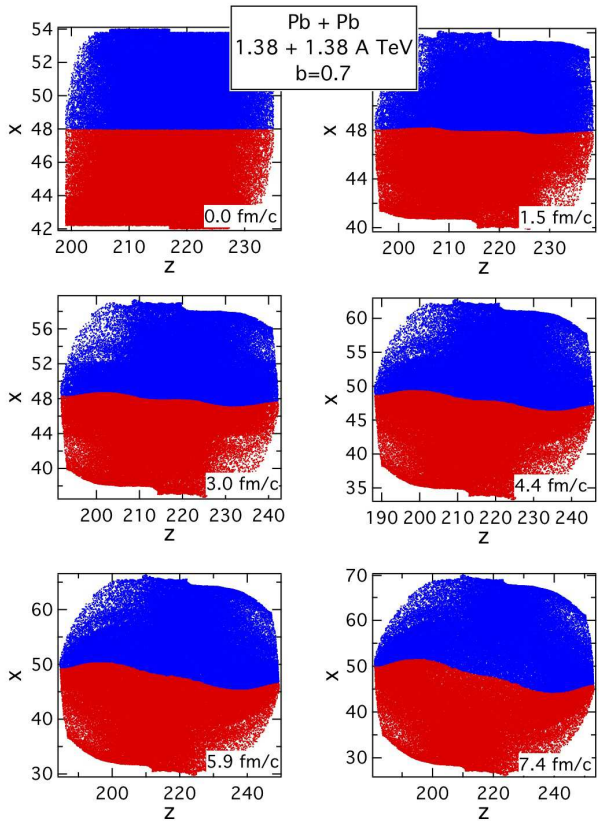}
\vskip -2mm
 \caption{
(color online)
Growth of the initial stage of Kelvin-Helmholtz instability 
in a $1.38A+1.38A$ TeV peripheral,
$b=0.7b_{\rm max}$, Pb+Pb collision in a relativistic
CFD simulation using the PIC-method. We see the positions 
of the marker particles (Lagrangian markers with
fixed baryon number content) in the reaction plane.
The calculation cells are $dx=dy=dz=0.4375$fm
and the time-step is 0.04233 fm/$c$
The number of randomly placed marker particles in each
fluid cell is $8^3$.
The axis-labels indicate the cell numbers in the $x$ and $z$
(beam) direction.
The initial development of a KH type instability is 
visible from $t=1.5$ up to $t=7.41$ fm/$c$ corresponding from 35 to 175 
calculation time steps).}
\vskip -5mm
 \label{F_growth}
\end{figure}

Our detailed studies indicate the development of an interesting 
phenomenon, namely the beginning of a physical instability. 
In peripheral collisions, in the transverse, [$y,z$] plane  a 
non-sinusoidal instability starts to develop.
We can visualize this by coloring  the markers in the projectile
(upper) side blue and the target (lower) side red. Initially
the dividing surface is a plane. As time proceeds, the markers 
(which indicate the location of conserved baryon charge) 
move, and the dividing surface becomes a wave, which resembles the
start of a Kelvin-Helmholtz (KH) instability as shown in 
Fig. \ref{F_growth} in the [$x,z$], reaction 
plane,  {\it i.e.}, $|y| \le 1$ cells ($|y| \le 0.7$ fm).

Initially, at 1.5 and 3.0 fm/$c$, we can see two shorter wavelength
KH instabilities, which then dissolve and are fed into a longer
wavelength instability.

The non-sinusoidal
behaviour of the instability at later times is not obvious as 
the development of instability and the spherical expansion
compete with each other.

 The density of the central zone  decreases
rapidly so that the matter ``freezes out'' and the fluid dynamical
description breaks down. As a consequence we do not see the 
vorticity sheet rolling up as in a fully fledged KH instability.
In more central collisions the dividing surface nearly 
remains a plane. 
The usual reason for the KH instability is a ``shear flow'',
where in a fluid layer there is a large velocity gradient. Thus, the 
origin  and the energy source of this phenomenon is in the initial 
configuration and the initial velocity distribution, which are 
correctly represented in our initial state model \cite{M2001-2}.

In our computational fluid dynamics (CFD) calculation
the initial state model  --based on longitudinally expanding 
flux tubes or {\it streaks} \cite{M2001-2}-- 
is used \cite{hydro1,hydro2}. In non-central collisions only
a part of the original nuclei interact. These are in the participant
zone where the streaks develop. Spectator nucleons on the two sides 
are not participating in the reaction. 

The participant streaks  are formed by the 
color charges arising from the projectile and target nuclei
after these have penetrated through each other. The chromo-electric
field, characterized by the {\it string tension}, slows down the
expansion of the ends of the streaks. Our FD initial state
is a configuration where the matter is stopped within each
streak, while streaks expand independently of each other. 
Thus, this model is applicable streak by streak and the
momentum of the streaks varies, especially for the peripheral streaks 
where the asymmetry between the projectile and target contribution
to the participant matter is the biggest. So, the streaks at the
projectile and target sides move in the beam ($z$) direction with 
substantial velocity difference. This generates the shear flow 
configuration.

The aspect ratio of matter in the reaction plane ($[x,z]$-plane)
becomes more elongated with increasing impact parameter, $b$,
as the height of the participant profile, $L$, becomes smaller, 
$L = (2 R - b)$, for nuclei of radius $R$, 
and the streaks are becoming longer due to the smaller effective
string tension \cite{M2001-2}. Thus, the aspect ratio for  
$b = 0.5 b_{\rm max}$ (where  $b_{\rm max} = 2 R$) is $[1:1.5]$, while for
$b = 0.7 b_{\rm max}$ it is $[1:3]$.  Of course this aspect ratio
depends on the initial state model, and some of these do not
take into account the longitudinal expansion before thermalization, and
even less the dependence of the expansion on the effective string tension.

In a heavy ion reaction the projectile edge of the participant domain
moves almost with the velocity of the projectile, $u$, while the target 
side moves with the target velocity, $- u$. At high energy this
difference provides considerable shear in the velocity fields.
At the same time in the initial state model \cite{M2001-2} the 
initial transverse velocity is zero for all fluid elements. 
A low $[x:z]$-profile makes it possible to develop a typical shear-flow
configuration and thus, there may be a possibility to form 
the initial stages of a KH instability.

\section{Physical Considerations}\label{Consi}

\subsection{Growth of the KH instability}
\vskip -3mm

The growth of a small initial KH instability in an idealized 
shear-flow configuration can be described in a rather simple way.
From ref. \cite{Drazin} sect. 3 it follows that
the shear flow starting from a small sinusoidal perturbation in
incompressible and inviscid flow, the perturbation will 
grow exponentially,  $\propto \exp(st)$, (\cite{Drazin} 3.15),
where 
$t$ is the  time and 
$s$ is proportional to the  wave number $k$,  
$$
s = k V      
$$
(\cite{Drazin} 3.28) 
where $\pm V$ is the characteristic velocity of the
upper/lower sheets which is somewhat less than the
projectile or target velocity.

Thus, the largest $k$ or shortest wavelength
will grow fastest.  Also, increasing the beam
energy (i.e. increasing $V$) will also lead to
increased development of turbulence!

For a Pb+Pb reaction, $R = 7$ fm, $b_{\rm max}= 14$ fm,
we study the impact parameters
$b = 0.5 \ {\rm or} \ 0.7 b_{\rm max}$. For these collisions the typical
transverse size of the initial shear flow is 
$$
L= (2R - b) = 7.0 \ {\rm or} \ 4.2\ {\rm fm}\ . 
$$
 The typical calculation
cell size is $dx=dy=dz=0.35$ fm. The beam directed,
longitudinal length of the initial state is
\begin{equation}
l_z = 10.5 \ {\rm or} \ 13.1\  {\rm fm} \ ,
\label{lz} 
\end{equation}
and the minimal wave number is 
$$
k = 2\pi/l_z = 0.598 \ {\rm or} \ 0.479\ {\rm fm}^{-1}
$$ 
respectively.

For the scaling analysis of instabilities we need the dimensionless 
numbers constructed from the typical length, $L$, and speed, $V$. 
The Reynolds number is $Re = VL/\nu$, where $\nu$ is the 
kinematic viscosity. So, for a peripheral heavy ion collision with
 impact parameter $b$, the 
characteristic length is $L$  and $V$ is the velocity of the 
top/bottom layer, $V=|u|$. In exactly central 
collisions  $u = 0$, while with increasing impact parameter,
$u = \pm 0.26, 0.34, 0.36, 0.39, 0.43, 0.42, 0.39\,c$ for
$b = 0.1, 0.2, 0.3, 0.4, 0.5, 0.6, 0.7\,b_{\rm max}$, respectively.
Notice that due to the geometry and the minimal string diameter
of 1 fm, the increase of this velocity does not reach the
beam velocity, so for typical peripheral collisions 
$V=|u| \approx 0.4$c.

In the simplest incompressible and inviscid flow approximation
the  amplitude of of the starting turbulence would double in
$2.90 \ {\rm or} \ 3.62$ fm/$c$ for $b = 0.5 \ {\rm or} \ 0.7 b_{\rm max}$.
The growth of instability is very fast in this approximation
and it increases with the 
beam energy (beam velocity) and with the wave number.
The typical reaction time in a heavy ion collision 
 exceeds the time needed to double
the amplitude of an initial instability. 

At the same time we also observe that at smaller impact
parameters the development of an instability is not seen
in our calculations, see Fig. \ref{F_growth-05}.
Thus, we have to conclude that the role of viscosity is decisive
as a large viscosity will decrease this growth rate and may
eliminate the possibility of the KH instability.

\subsection{Formation of critical length KH instability}

While perturbations with larger wave number (shorter wavelengths)
may grow faster, there is a critical minimal wavelength 
beyond which the
perturbation is stable and able to grow. Smaller wavelength 
perturbations tend to decay into random thermal fluctuations. 
This situation is analogous to the
phase transition dynamics via homogeneous nucleation where
the formation of critical size bubbles or droplets is required
to start the phase transition \cite{CK92}.

This aspect of turbulence formation was first discussed by
Kolmogorov  \cite{Kolmogorov}
for flow in the ``inertial range'' where the
effects of viscosity are still negligible. 
These considerations are applicable until the
viscosity does not have a significant effect
on the formation of vortices.
The minimal stable wavelength for a starting
instability is given by the Kolmogorov length scale,
which is the smallest scale of turbulence. 
Dominant and increasing viscosity results in increasing
critical vortex size, $ \lambda_{\rm Kol} $.
Smaller perturbations are unstable.  

The average rate of energy dissipation, $\epsilon$, 
per unit mass and unit time, is associated with the decay of
an eddy of size {\it l} and characteristic speed, $v_l$ into two 
smaller ones, in time $t_l = l / v_l$. It follows then that 
$\epsilon \sim v_l^2 / t_l = v_l^3 / l$. The decay (or formation)
time of the small size eddy, $t_l$, can be compared to the viscous
diffusion time of a perturbation of size $l$, which is
$t_l^{dis.} = l^2/\nu$. Equating the two estimated characteristic
times provides the minimal, Kolmogorov length, $\lambda_{\rm Kol}$:
\begin{equation}
\lambda_{\rm Kol} = [ \nu^3 / \epsilon ]^{1/4}.
\label{Kolmo}
\end{equation}
Here $\nu$ is the 
kinematic viscosity of the fluid, 
$\nu = \eta / \rho = \eta / ( n m_B)$,
where $\eta$ is the shear viscosity, $\rho$ is the mass
density, $n$ is the baryon charge density and $m_B$ is the 
characteristic mass falling on unit net baryon charge 
in QGP. 

   Even if the average rate of energy dissipation, $\epsilon$,
is proportional to the viscosity, this dependence is linear,
so the critical vortex size is still increasing with increasing
viscosity. The key question is: can a critical size vortex
be formed in a heavy ion collision? Lower viscosity and higher
energy or energy dissipation may enable the formation of a
critical size or larger vortex.

Kolmogorov's theory also provides an energy distribution spectrum
for small vortices or whirls in nearly perfect fluids. The energy
density spectrum in terms of the scale of the vortices, $\lambda$,
is proportional
to $\lambda^{-1/3}$, {\it i.e.}, it is lower for larger vortices.
Large vortices may generate smaller ones, until we reach the 
viscous limit, $\lambda_{\rm Kol}$, where the vortices are becoming 
just thermal fluctuations.

In non-relativistic flow the mass flow is identical
with the flow of massive particles while the flow energy 
and the thermal energy are negligible compared to the rest
mass of the fluid. At ultra-relativistic energies this is not the case
and the separation of flow (inertial) energy and the random
thermal energy is not a trivial question. In order to
follow the classical concepts turbulence and its development
we will follow Eckart's definition of flow velocity, where 
flow is bound to the conserved net baryon charge and we will assume
that the flow of the average of all quarks can be characterized
by this velocity. The observed constituent quark number
scaling of collective flow observables for different hadrons
supports this approach.
  
We focus on the description of the initial stages of
the development of turbulent instability in the central zones
of the collision at a period just after the formation
of the locally equilibrated FD initial state. 
We will estimate the corresponding collective mass of the
partonic matter in QGP per unit baryon charge, $m_B$, at this
stage of the reaction. Before the collision at the LHC each nucleon has
1.38 TeV energy.  Local equilibration is reached when, for 
$b=0.7 b_{\rm max}$ impact parameter,
in the first 3 fm/$c$ time after initial equilibration the average
temperature is $T \approx 400 \div 600$ MeV, the average entropy
density is $s \approx 150 \div 440$ fm$^{-3}$, the average baryon
charge density is $n = 0.1 \div 0.16$ fm$^{-3}$ and the average
internal energy is 
$T (s / n) = T \sigma = e/n \approx 1.1 \div 1.3$ TeV/nucl., where
$\sigma$ is the specific entropy per unit baryon charge. 
At the initial stages of QGP flow 
the remaining energy is shared between collective flow energy and
particle mass of all constituents, plus a
smaller amount may go to pre-equilibrium emission of high energy
particles. Thus, the effective mass can be estimated 
as  $m_B \approx 100$ GeV per unit baryon charge.

By late stages of the 
reaction the kinetic energy of the flow increases substantially, while
the dissipation increases the thermal energy by about 
$4 \div 6$\% \cite{Horvat}.
The total hadron multiplicity increases by about an order of magnitude,
so the effective mass per net baryon charge is of the order of 
$m_B \approx 10$ GeV.
We are interested in the case of initial QGP in local equilibrium
when the KH instability could start.
 
The shear viscosity is temperature dependent with a sharp
minimum at the critical point of the phase transition between
hadronic matter and QGP \cite{CKM}, and reaching unity at
the initial hot, compressed QGP at about $4 T_c$: 
$$
\frac{\eta}{s} \approx 1\ \div\ 2\ \hbar ,
$$
which for the initial QGP gives
$$
\eta = s \hbar \approx  30 \div 158\ \mbox{GeV/(fm$^2$\,c)},
$$
and about 10 times less for the minimal viscosity.
Considering that the initial effective mass density may vary between,
$\rho = n m_B \approx 10 \div 16$ GeV/(fm$^3$\,c$^2)$, the 
corresponding kinematic viscosity is
$$
\nu = \frac{\eta}{\rho} = 2.5 \ \div\ 16\ \ {\rm fm}\,c \ ,
$$
and about 10 times less for the minimal viscosity.

The corresponding Reynolds number is 
$Re = 0.3\ - 1$ 
(for ``$\eta/s = 1$''),
and  if we choose the minimal viscosity 
(``$\eta/s = 0.1$''), 
then $Re = 3 - 10$.  
These are small $Re$ values for turbulent flow in general, 
but the KH instability can also appear for small $Re$ \cite{Drazin}.
\begin{figure}[h]
 \centering
 \includegraphics[width=3.4in]{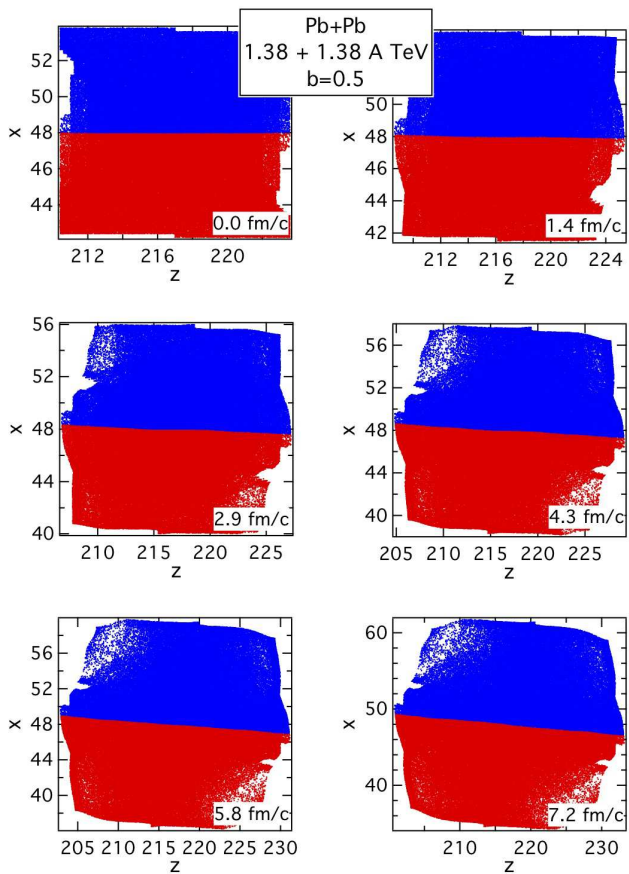}
 \caption{
(color online)
Time evolution of the flow 
in $1.38A+1.38A$ TeV peripheral,
$b=0.5b_{\rm max}$ Pb+Pb collision.
The calculation cells are $dx=dy=dz=0.585$ fm 
and the time-step is 0.08466 fm/$c$.
The number of randomly placed marker particles in each
fluid cell is $8^3$.
In contrast to Fig. \ref{F_growth} 
the KH instability does not develop during the initial
 $\sim 8$ fm/$c$ time, due to the
increased numerical viscosity and the similar length and height
of the initial state. Thus the sides of the fluid provide
a stiffer formation and the system rather rotates as a solid body, 
instead of forming an expanding turbulent, rotating shell.
Due to angular momentum conservation the rotation slows 
down as the system expands.}
 \label{F_growth-05}
\end{figure}

If we have a perfect fluid the flow is
adiabatic, there is no dissipation, so
$\lambda_{\rm Kol} \sim (0/0)$ !
The specific dissipated flow energy to heat is
\begin{equation}
\epsilon = \dot{e}/\rho\ \ \propto\ \ T \dot{\sigma}/\rho\ \ 
\propto\ \ \nu ,
\label{Eps1} 
\end{equation}
where 
the dot indicates the proper-time derivative, 
$\dot{e} \equiv \partial_{t} e$ 
is the change of energy density with time,
$T$ is the temperature and  $\dot{\sigma}$ is the
proper time derivative of the specific entropy density,
$\sigma$.
Thus if $\nu \longrightarrow 0$ then
$
\lambda_{\rm Kol} \longrightarrow 0  .
$
So, the minimal size will grow from zero if the viscosity
grows. With finite viscosity one can have a large minimal size, 
so that the turbulence can not develop within the given length of the
system. As in the final expanding stages of the QGP fluid the viscosity
increases \cite{CKM}, the minimal eddy size, $\lambda_{\rm Kol}$, 
will be larger, so initial smaller length instabilities will
disappear.

A good example for the formation of a minimal size eddy can be observed 
in a two-component Fermi-gas (e.g. Li-6), which forms a super-fluid at 
low temperatures in a rotating magnetic trap \cite{HW201}. 
If we reach a
limiting rotation frequency, a small eddy may develop in the central region
of the cylindrical trap when the energy of a small critical size eddy
will be sufficient to balance the viscous dissipation, and this eddy may
become stable. One also needs a given viscosity or scattering length
to form this eddy in the middle, thus the eddy first appears at a
given finite scattering length! See Fig. 14 of ref. \cite{HW201}a. 
The minimal KH instability  has similar minimal size behaviour, although
the shear-flow geometry is different; one needs a minimal torque
from the boundary condition and a minimal viscosity to form a
critical eddy.

Let us make a very simplified estimate for the size of the 
smallest possible eddy in a heavy ion collision.
For finite shear viscosity
the energy dissipation per unit mass and unit time,
$
\epsilon = \dot{e}/\rho 
$,
depends on the viscosity as well as 
the flow pattern.
The characteristic shear $V/L$ depends on the impact parameter,
so that $(V/L)^2 = 0.0038 \div 0.0086 $ (c/fm)$^2$ for 
$b=0.5 \div 0.7 b_{\rm max}$, respectively.
For the ideal shear flow the dissipated energy 
with the minimal viscosity (``$\eta/s = 0.1$'')
is (see \cite{LL} sect. 16 or
\cite{Davidson} sect. 1.6):
\begin{eqnarray}
\dot{e}  &\approx&
\eta\, \left(   \frac{V}{L} \right)^2 = 
\label{edot} \\ 
 &=& \left\{
\begin{array}{ll}
   11 \div \ 26\ {\rm MeV}\,c/\,{\rm fm}^4\ \mbox{for\ $b=0.5 b_{\rm max}$}  \\
   61 \div  138\ {\rm MeV}\,c/\,{\rm fm}^4\ \mbox{for\ $b=0.7 b_{\rm max}$}. 
\end{array}
\right. 
\nonumber
\end{eqnarray}
Then the energy dissipation, with an estimated average
unit mass density of $\rho = 13 $ GeV/(fm$^3$ c$^2)$, gives the 
rate of energy dissipation
$
\epsilon = \dot{e}/\rho 
$
and thus the Kolmogorov length is
\begin{equation}
\lambda_{\rm Kol} = \left\{
\begin{array}{ll}
   2.1 \ \div \ 5.4 \ {\rm fm}\ \mbox{for\ $b=0.5 b_{\rm max}$}  \\
   1.4 \ \div \ 3.6 \ {\rm fm}\ \mbox{for\ $b=0.7 b_{\rm max}$}  
\end{array}
\right.
\label{length}
\end{equation}
In peripheral heavy ion collisions
the KH instability can develop only if
\begin{equation}
l_z > \lambda_{\rm Kol} \ .
\label{Kol} 
\end{equation} 
Thus, comparing the above values of the Kolmogorov length
scale to the length of the initial state in the beam direction,
eq. (\ref{lz}), we can see that at $b=0.7 b_{\rm max}$ we may have
a possibility to initiate a KH type instability in a heavy
ion collision, while at smaller impact parameters this
possibility is marginal. This is enhanced by the fact that
viscosity and the Kolmogorov length increases with expansion,
so the time-slot where the KH instability may develop
is reduced for more central collisions.

Based on eqs. (\ref{Kolmo},\ref{edot}) we see that in our
situation the Kolmogorov length is proportional with the square
of viscosity, 
$
\lambda_{\rm Kol} \propto \eta^2 \ ,
$
so if the viscosity doubles, the Kolmogorov length, eq. (\ref{length}), 
increases by a factor of four, so it can reach 20 fm, which exceeds the
longitudinal system size, and hinders the development of
the KH instability up to about 8 fm/$c$. This is illustrated in Fig.
\ref{F_growth-05} with the increased numerical viscosity.
The change of marker particle resolution does not influence the
disappearance of the KH instability. 

\begin{figure}[h]
 \centering
 \includegraphics[width=3.4in]{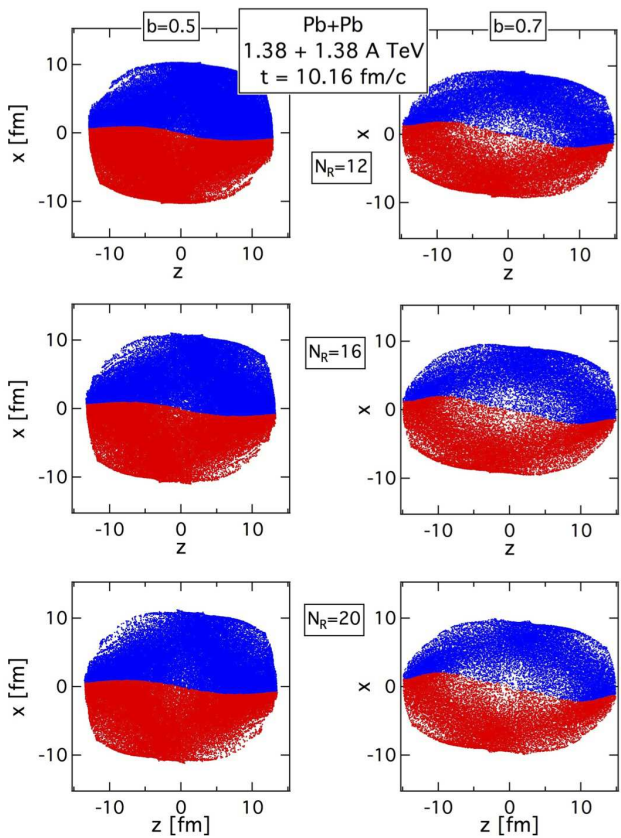}
 \caption{
(color online)
Comparison of the flow pattern in the reaction plane in 
$1.38A+1.38A$ TeV peripheral,  Pb+Pb collisions at two
impact parameters: $b=0.5b_{\rm max}$ (left column) and
$b=0.7b_{\rm max}$ (right column),  at a late stage of 10.16 fm/$c$
(240 time-steps). By this time an instability wave is also noticable 
at $b=0.5b_{\rm max}$.
The resolution increases from the top to the bottom as
$dx= 0.585, \ 0.4375, \ 0.35$ fm (i.e. $N_R=R_{\rm Pb}/dx=12, \ 16, \ 20$).
The number of marker particles were
$8^3$, $6^3$ and $5^3$ for these resolutions, so that each of 
the marker particles carried about
the same amount of baryon charge: 5.61, 5.61, 4.97 $\times 10^{-5}$.
For the impact parameter  $b=0.5b_{\rm max}$ independently of
the resolution and numerical viscosity the rotation of the
dividing plane is $2$ degrees, and the amplitude of the KH
instability wave is not bigger than 0.35 fm. 
For the impact parameter  $b=0.7b_{\rm max}$ the rotation of the
dividing plane is $3.75$ degrees for all resolutions, but 
the amplitude of the KH
instability wave is increasing with increasing resolution
(decreasing numerical viscosity) as: 0.7, 0.9 and 1.1 fm.
}
\label{F_resolution}
\end{figure}

The characteristic geometry of the KH instability
(approximately two planes close to each other)
require that  $L < R$ or $b \ge R$,
and so $b$ may vary  in the interval between
$0.5 b_{\rm max}$ and $0.8 b_{\rm max}$ where  $b_{\rm max} = 2 R $.
At larger impact parameters $L$ becomes too small and
the general applicability of a continuum approach
becomes questionable.
So, this parameter cannot vary too much, and
condition (\ref{Kol}) requires to have a viscosity
smaller than a limiting value,
$$
\nu_c \simeq 5 \ {\rm fm}\,c,
$$
which is satisfied by the low viscosity QGP, even at
higher than critical temperatures.

\section{Fluid Dynamical Model Predictions}

We have performed CFD simulations with the PIC solution
method \cite{hydro1,hydro2} where the equations
of relativistic FD were solved for a perfect quark-gluon fluid. 
At the same time the numerical method, due to the finite grid 
resolution, led to dissipation and thus to entropy production,
which has been analysed \cite{Horvat}. From the entropy production
we could determine the corresponding ``numerical viscosity'',
and this was approximately the same as the estimated, low viscosity 
of the quark gluon plasma. To avoid double counting, {\it i.e.}, over 
counting of viscous dissipation, we did not add additional viscous
terms to our CFD model simulations \footnote{
In numerical, CFD simulations of instability, it is important
to study the dissipative effects of both the numerical viscosity
and the physical viscosity. A finite grid resolution in the CFD
solutions leads to the absorption of the shorter wavelength and high
frequency fluctuations, and the energy of these fluctuations is
converted into heat. 
There exist solution methods, with finite computational grid 
resolution, which enforce entropy conservation for perfect fluids.
These solutions then appear to be perfect adiabatic fluid 
flow solutions. However, this is misleading, such a numerical 
solution still absorbs high frequency small wavelength perturbations, 
while the energy of these is converted into large wavelength 
fluctuations. Thus, such methods may result in misleading results.  
}.

Our numerical model predictions confirm the previously presented 
physical conclusions, and these show a developing instability
which is visible  at $b=0.7 b_{\rm max}$ (Fig. \ref{F_growth}),
but at the smaller impact parameter the KH instability is
weak and does not change with increasing resolution, see Fig. 
\ref{F_resolution}.  Increasing the resolution by 67\%  
at $b=0.7 b_{\rm max}$ increases the amplitude of the KH instability wave 
by 57\%.  The final KH instability amplitude (1.1 fm) is 6\% of the 
final profile height (and 16\% of the initial one).

In the case of heavy ion collisions  the special 
geometry, {\it i.e.}, the shape of the participant zone, is also hindering 
the development of the instability because the more extended
side-walls at smaller impact parameters have a stabilising effect 
against the KH instability.

We can observe in our calculations that  
the initial sinusoidal wave shape will become
asymmetric in standard plane shear flow (see \cite{Drazin} Fig. 3.3 or  
\cite{Krasny} Fig. 2), especially at points of accumulating vorticity.

The PIC method has a particular advantage in studying the KH instability.
The numerical viscosity is set by choosing a calculation grid size,
which provides the estimated small dissipation of the viscous QGP fluid. 
At the same time the PIC method has a large number of
marker particles in each Eulerian fluid cell. Their number can vary, and can be orders of magnitude higher then 
the number of Eulerian fluid cells forming the fixed calculation grid.
Thus the motion of the marker-particles provides a fine resolution and
can follow the dynamics of the flow more accurately. In this method
the marker particles provide an accurate tracing of the initial development
of the KH instability.

In connection with our CFD solution one has to mention that the
{\it numerical viscosity} of our model calculation is small,
$\eta/s = 0.1$, based on the small Eulerian cells. In addition 
the number of initial marker particles per normal density cell was 
changed from $3^3 = 27$ to $9^3 = 729$ so that the higher Lagrangian
resolution allows for a more accurate description of the instability. 

In case of heavy ion reactions the flow is not stationary
and the shear flow geometry is only present in the initial
state.  Later, due to the large pressure of QGP the plasma
explodes and expands radially in a way that the final flow pattern
at freeze out is close to spherical albeit somewhat elongated longitudinally 
and in the reaction plane ($\pm x$-direction) due to the dominant
elliptic flow.

\subsection{CFD Results}

In heavy ion reactions
the radial expansion modifies the dynamics of the development
of the KH instability, but in case of low viscosity this is a 
strong instability and its initial signs can be clearly recognized 
in CFD calculations if both the viscosity and the numerical viscosity
are sufficiently small.  
\begin{figure}[h]
 \centering
 \includegraphics[width=3.4in]{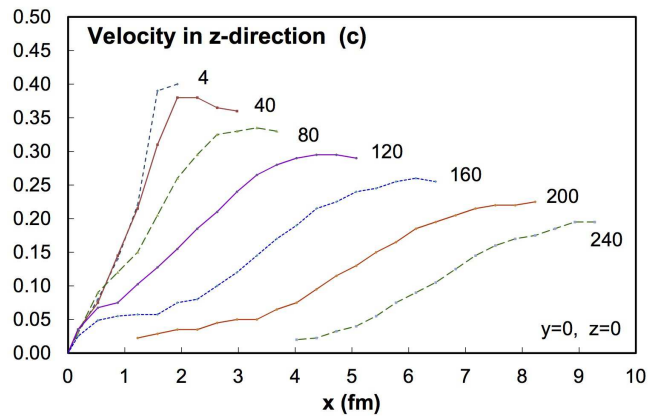}
 \caption{
(color online)
The velocity profile in the beam direction  as a function of the
$x$-coordinate, at different times, 4, 40, 80, 120, 160, 200 and 240 
time-steps of 0.04233 fm/$c$ for $b=0.7 b_{\rm max}$, cell size $dx=0.35$ fm
and $7^3$ marker particles per fluid cell.  The velocity is plotted in the 
reaction plane, (at $y=0$ and $z=0$), and it is antisymmetric
in the $\pm x$-direction, but only the upper $x>0$ half of the
reaction plane is plotted. In the CFD calculation the mirror symmetry
is exact. }
 \label{F_VsProf}
\end{figure}
Looking at Fig. \ref{F_growth}
initially
   at $N_{\rm cyc}$=0, the fluctuation is not visible, although
the randomly placed markers include a possibility
for fluctuations (the length of the system
is 35 cells).
   At 1.5 fm/$c$ ($N_{\rm cyc}$=35), there appear two semi-sinusoidal waves.
The length of the system is 45
cells. The amplitude of turbulence is $\approx$ 1 cell.
   At 3.0 fm/$c$ ($N_{\rm cyc}$=70) the length of the system is $\approx$ 50
cells. The central  perturbation of 15 cell wavelength 
weakening while the outside part grows. 
The amplitude of turbulence is $\approx$ 3
cells. 
  At 4.4 fm/$c$ ($N_{\rm cyc}$=105) the middle middle
perturbation wave is weakening further, and the 
amplitude of the turbulence is $\approx$ 4 cells.
   By 5.9 - 7.4 fm/$c$ ($N_{\rm cyc}$=140 $\div$ 175) the 
middle, short wavelength perturbation is
hardly visible.  The amplitude of the turbulence has
reached 6 $\div$ 10 cells.

Due to the fact that the radial expansion and 
the shear flow are superimposed upon each other
the growth rate and the wave shape of the developing
turbulence are not identifiable fully as in (quasi-) 
stationary flow.
The initial shorter wavelength perturbations become unstable
and disappear, while the longest one grows due to the
radial expansion. This can be attributed to the
fact that the Kolmogorov minimal size is increasing
faster than the expansion, thus the short perturbations are
becoming unstable while the largest one survives.

The phenomenon enables us to draw some quantitative 
consequences from the physical viscosity of QGP on
qualitative differences in the flow pattern.
  The smaller central perturbation which disappears 
during expansion is not detectable.
  The larger one develops if the available length
of the system exceeds the Kolmogorov length scale.
Then the shortest of these possible perturbations 
will grow fastest and will lead to enhanced and
observable ``rotation''. Increased beam energy leads
to increasing $V$, which leads to an exponential
increase in the growth of the instability, 
much more than just the linearly increased angular
momentum would cause!

As the estimates of the previous section indicate,
the development of the KH instability is critically dependent
on the flow configuration, and just as the numerical
estimates for $\lambda_{\rm Kol}$ indicate the 
CFD results for impact parameter $b = 0.5 b_{\rm max}$ also
show that the KH instability is weak and it does not develop 
with increasing resolution for
more central collisions, see Fig. \ref{F_growth-05}. 

The dynamics in the CFD model indicates that the
KH instability may start to develop in ultra-relativistic
heavy ion collisions. The numerical viscosity 
in the calculation is about the same as the 
conjectured minimal viscosity \cite{Son} as
discussed in ref. \cite{Horvat}.

The analysis in the previous section assumed an initial
viscosity based on ref. \cite{CKM} which was around 
the minimal viscosity and it was sufficient
to develop the critical size KH instability. An order of 
magnitude larger viscosity would not
be able to create a sufficiently small initial
perturbation.

\subsection{Instability estimates for viscous fluids}\label{VF}

In shear flow one of the  necessary conditions for
the start a KH instability is that there should be an inflection 
point in the basic velocity profile $u_z(x)$, or in other words
$u''_z(x)$ must change sign at least once \cite{Drazin} Sect. 8.
The initial condition \cite{M2001-2} and the subsequent flow
satisfy this requirement.

As Fig. \ref{F_VsProf} shows, at time-step 4 and 40 the condition is 
satisfied in the center ($x=0$), later at steps 80, 120 and 160
a second wave develops, however the distance of the nodes  is less than
$\lambda_{\rm Kol}$, so considering the viscous limits these 
secondary waves are not realizable physically. For a single
wave of KH instability the situation is established, and persists
up to time-step 120.  By time-step 160 the central density drops
considerably and apart from short length
fluctuations the required condition is not satisfied.

\begin{figure}[h]
 \centering
 \includegraphics[width=3.4in]{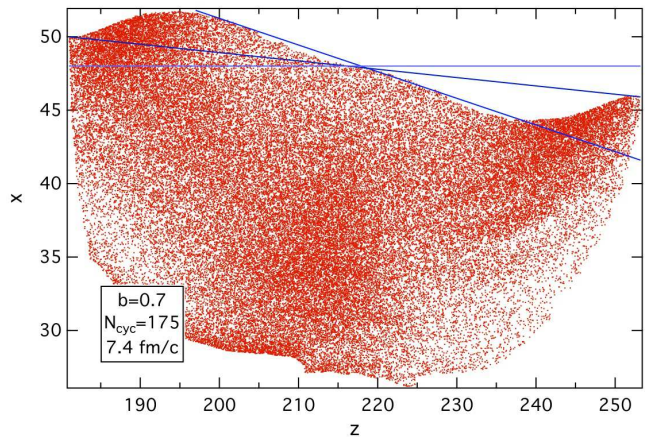}
\caption{
(color online)
The detailed view of the marker particle positions in the lower half 
of the initial state markers after 175 time-steps. A $1.38A+1.38A$ TeV 
energy  Pb+Pb peripheral collision is shown, at $b=0.7\ b_{\rm max}$ impact
parameter with $7^3=343$ markers per initial, normal density fluid cell
resolution. The lines
across the collision center point indicate the initial dividing axis, the
change of this axis due to rotation and the additional change of rotation
arising from the start-up of a Kelvin-Helmholtz type of instability.
This additional effect more than doubles the rotation.
In this calculation the cell size is $dx=dy=dz=0.35$ fm,
with a total number of 1814814 marker particles. 
 }
 \label{F_Gr-b7-343}
\end{figure}

This is due to the fact that in the discussed heavy ion collision,
we do not have a stationary boundary condition, and the spherical
expansion competes with the growth of KH instability. At later 
stages, time-step 160 and later, the growth speed, $k\, V$ decreases,
as both the velocity and the wave number decrease, while the
expansion speed is increasing. Thus at time-step 160 the flow
goes over to a combined approximately radial expansion and rotation.
On the other hand by this time the system is close to freeze out and
hadronization.  At this time the matter is dilute and weakly interacting,
so the local fluid dynamical equilibrium cannot be maintained
and the FD description will not be applicable.

The central zones at late stages have low density and low pressure,
which resembles a cavitation bubble in the QGP liquid and this
bubble will contain the condensed hadrons. Due to the expansion
the bubble will not re-collapse as usual in cavitation phenomena
in classical fluid flow; instead the surrounding QGP fluid will
break up into peaces and will also hadronize and freeze out 
simultaneously.

\section{Conclusions}\label{Conc}

Based on
theoretical estimates and on CFD model calculations, one 
should explore the possibility  at LHC
energies of a
KH instability developing in peripheral, 
$b=0.6-0.8 b_{\rm max}$ Pb+Pb collisions. The formation
of the instability may take place up to about 5 fm/$c$, beyond which the
radial expansion becomes dominant although the system still rotates.

The KH instability is rather sensitive to the value of viscosity, so
it is a perfect tool to measure the viscosity of QGP. The
rotation of the weak anti-flow peak  to forward angles was predicted
earlier \cite{hydro2}. The rotation of the $v_1$-peak to forward angles
depends sensitively on the balance between the speed of radial 
expansion/explosion and the initial angular momentum 
(which increases with increasing beam energy). 
If the radial expansion is stronger than estimated \cite{hydro2}
then the peak may remain an "antiflow" peak and the KH instability
would destructively interfere with this peak. 
If the peak has rotated to forward angles as predicted in \cite{hydro2}
and used also in these calculations, the KH instability increases the 
rotation and
it converts a larger part of beam energy into rotation than it would
happen in a simple solid body type of rotation. See Fig. \ref{F_Gr-b7-343}.

At cycle 160 ($t=6.77$ fm/$c$) using the method of 
ref. \cite{hydro2} primary $v_1(y)$ values for massless pions
were evaluated for two different impact parameters, $b= 0.5, 0.7 b_{\rm max}$,
and for different grid resolutions. The number of marker particles were
chosen so that the number of marker particles per baryon charge was
about the same, 
20120 for cell size $dx=0.35$ fm  and 
17825 for cell size $dx=0.585$ fm.  
For the Pb+Pb reaction at
$b=0.5b_{\rm max}$ and $dx=0.585$ fm resolution the directed flow
peak was at the rapidity bin, $y=0.45$ with a peak value of 
$v_1(0.45) = 0.177$. This was taken to be 100\%. By decreasing the 
cell size to  $dx=0.35$ fm the peak value increased to 
$v_1(0.45) = 0.200$, i.e. by 13\%.
The changes of
the directed flow peak values are shown in Table I.

\begin{tabular}{ccrrrr}
\\
\multicolumn{6}{c}{TABLE I. Change of Directed Flow}  \\
\hline\hline
y    &                 & Pb+Pb & Pb+Pb & Pb+Pb & Pb+Pb  \\
     &$b/b_{\rm max}$      &  0.5  &  0.5  &  0.7  &  0.7   \\   
     &$dx, dy, dz$ [fm]& 0.585 & 0.350 & 0.585 & 0.350  \\
\hline
0.35 & $v_1(y)$ [\%]   &  -3   &  10   &  25   &  47    \\
0.45 & $v_1(y)$ [\%]   &       &  13   &  32   &  53    \\
0.55 & $v_1(y)$ [\%]   &  -3   &   9   &  20   &  50    \\
\hline\hline
\\
\end{tabular}

For $b=0.5 b_{\rm max}$ the KH instability on the $v_1$-peak is weak, 13\%,
which can partly be attributed to the decreased viscous dissipation.   
At $b=0.7 b_{\rm max}$, the increase is 21\%, it is significantly stronger.
These primary data of of course are reduced by random initial state
dissipation, as discussed in ref. \cite{hydro2}. The position of the 
peak in rapidity is hardly changing with higher grid resolution,
for $b=0.5b_{\rm max}$ it moves from $y=0.485$ to $y=0.46$, while 
for $b=0.7b_{\rm max}$ it moves from $y=0.475$ to $y=0.46$.

Recently if was pointed out \cite{Urs} that due to random initial
fluctuations turbulence may show up and even grow in the transverse,
$[x,y]$ plane. The energy of the growth is provided by absorbing 
small, higher wave number perturbations. We also observed this effect
(see Fig. \ref{F_growth}). In that work an alternative 
detection method is suggested via measuring two particle correlations, 
which may also be used to detect the KH instability in the
reaction plane.

Although, the predicted rotation effect is not easily detectable
due to initial state fluctuations, the KH instability 
enhances the flow and changes its pattern in peripheral collisions. 
The present developments suggest that the global collective
$v_1$ flow can be disentangled from random fluctuations. This is 
necessary to measure the global collective flow in peripheral
collisions. The opposite, the separation of the flow originating
from random initial state fluctuations is done successfully recently
\cite{ALICE-Head-On} for selected central collisions.

The KH instability is very sensitive to the magnitude of the viscosity.
Thus if this 
research is successful the analysis of global collective $v_1$ flow as a 
function of beam energy and impact parameter may provide a precision
measurement of viscosity and its variation.

\section*{Acknowledgements}\label{Ack}

We thank Martin Greiner, Peter Van and Pawel J. Kosinski for
valuable discussions. L.P. Csernai and D.D. Strottman thank
for the hospitality  of Dirk Rischke, of the Frankfurt Institute 
for Advanced Studies and of the Alexander von Humboldt Foundation.


\end{document}